\documentstyle[aps,prl,twocolumn,epsfig]{revtex}

\begin{document}
\draft \twocolumn [\hsize\textwidth\columnwidth\hsize\csname
@twocolumnfalse\endcsname \draft
\title{ Thermodynamics of Two Dimensional Magnetic Nanoparticles. \\}
\author{P. Vargas$^{1,2}$ , D. Altbir$^{3}$, M. Knobel $^{4}$ and D. Laroze$^{5}$}
\address{$^1$Departamento de F\'{\i}sica, Universidad T\'{e}cnica
Federico Santa Mar\'{\i}a, Valpara\'{\i}so, Chile\\
$^2$ Max-Planck Institut f\"ur Festk\"orperphysik, D-70569
Stuttgart, Germany.\\
 $^3$
Departamento de F\'{\i}sica, Universidad de Santiago de Chile,
Av. Ecuador 3493, Santiago, Chile\\
$^{4}$ Instituto de F\'{\i}sica Gleb Wataghin, Universidad
Estadual de Campinas (UNICAMP),Caixa Postal 6165,Campinas
13083-970 ,SP, Brasil.\\
 $^{5}$ Instituto de F\'{\i}sica,
Universidad Cat\'olica de Valpara\'{\i}so, Valpara\'{\i}so,
Chile.\\ }

%\date{\today}
\maketitle

\begin{abstract}
\noindent A two dimensional magnetic particle in the presence of an
external
magnetic field is studied. Equilibrium
thermodynamical properties are derived by evaluating analytically
the partition function. When the external field is applied
perpendicular to the anisotropy axis the system exhibits a second
order phase transition with order parameter being the
magnetization parallel to the field. In this case the system is
isomorph to a mechanical system consisting in a particle moving
without friction in a circle rotating about its vertical diameter.
Contrary to a paramagnetic particle, equilibrium magnetization
shows a maximum at finite temperature. We also show that uniaxial
anisotropy in a system of noninteracting particles can be
missinterpreted as a ferromagnetic or antiferromagnetic coupling
among the magnetic particles depending on the angle between
anisotropy axis and magnetic field.
\end{abstract}

PACS numbers: 75.10.-b, 75.30.Gw, 75.50.Tt

\vspace{0.5cm}

]

\narrowtext

The remarkable development of experimental techniques grants
increasingly more access to the hitherto mysterious nanometric
world. In the particular case of magnetism, a recent experimental
study has succeeded to investigate single clusters of around 3 nm
diameter, using a special micro-SQUID device \cite{Jamet01}. Also,
a wide variety of magnetic microscopies
\cite{DanDahlberg99,Wernsdorfer00,Rothman01} are under constant
development to measure the magnetization process of single
clusters in the nanometer range. In this context, the classical
Stoner-Wohlfarth (SW) \cite{stoner48} model has been frequently
revisited, to test wether or not such simplistic uniform
rotation model really provides a realistic picture of the
magnetization reversal of a nanoparticle
\cite{Bonet99,Wernsdorfer97}. The SW model does not take
explicitly into account thermal fluctuations, and therefore it is
strictly valid for very low temperatures or very strong
anisotropies \cite{garciaotero98,Dimitrov96}. On the other hand,
when the anisotropies are rather small and/or the temperatures are
sufficiently high, the so-called superparamagnetic approach must
be employed. In such case, N\'{e}el \cite{Neel49} and Brown
\cite{Brown63} introduced the idea that the decay toward a thermal
equilibrium state of the magnetic moment of a single-domain
ferromagnetic nanoparticle is mediated by thermal fluctuations,
following a simple Arrenhius switching probability
\cite{Wernsdorfer97,Igarashi00}. Also in this case, further
improvements of the model have been introduced year after year in
order to include additional terms which are usually present in
real samples, such as anisotropy
\cite{Cregg99,Respaud99,Pfannes00} and dipolar interaction among
magnetic nanograins \cite{Allia99,Anderson97,Allia01,Denardin01}.

In this letter, we present a mechanical analog to a magnetic
moment in the presence  of an external field perpendicular to its
uniaxial anisotropy axis.  This particular system exhibits a
second order phase transition as a function of the external field,
where the order parameter is the magnetization. Its
thermodynamics is also analytically solved showing the behavior of
the magnetization and susceptibility as a function of temperature.
Finally we demonstrate that uniaxial anisotropy in an ensemble of
non interacting SW particles can be missinterpreted as a ferro or
antiferromagnetic coupling among  the particles.

Let us consider a single magnetic domain particle with all its
atomic moments rotating coherently on a plane in the presence of
an external magnetic field. The particle is characterized by a
constant absolute value of the magnetization $m = m_s V$, $V$
being the volume of the particle and $m_s$ its saturation
magnetization. Well below the Curie temperature of the specific
magnetic element, $m_s$ is independent of particle volume and
temperature. The energy of this nanoparticle has three
contributions: anisotropy energy, either due to shape, stress or
crystalline structure of the particle; Zeeman energy, owing to the
presence of an external field, $\vec{H}$, and the magnetic dipolar
interaction among the atoms composing the grain,
and  which will not be considered in the present work. Under these
conditions the energy reads:

\begin{equation}
E\; = \;  -\vec{m}\cdot \vec{H}-\kappa V \left(
\frac{\vec{m}_i\cdot \hat{n}}{m}\right)^2  \;\;,
\end{equation}

\noindent where $\kappa$ is the anisotropy constant and the unit
vector $\hat{n}$ represents the easy magnetization direction. We
will initially focus on the particular case in which the external
field is perpendicular to the anisotropy axis. Then, the  energy
of the single particle can be written as:

\[
E= \; - \;  m H \cos{\theta} - \frac{1}{2} m H_a \sin^2\theta \;
\; \;.
\]

\noindent where we have defined the anisotropy field $H_a$ such
that $\kappa V = \frac {1}{2} m H_a$ and $\theta$ represents the angle
between magnetization and external field.

A more suggestive form is obtained by making the energy dimensionless
as follows:

 \[
 {\cal{ E}} = \frac{E}{m H} = -\cos{\theta} -
 \frac{1}{2}\frac{H_a}{H}
\sin^2{\theta} \; \;.
\]

Now, let us consider a rigid circular ring of radius $R$
rotating along its vertical diameter at angular frequency $\omega$
with a frictionless particle of mass $m$ free to move on the ring, as
illustrated in Fig. 1.
The particle is under gravity, $m \vec{g}$, and normal force
$\vec{N}$ due to the ring, which is rotating, and therefore the
mass describes a circular orbit due to the centripetal
acceleration originated from the horizontal component of
$\vec{N}$. By choosing the zero level of gravitational energy at
the center of the ring, the Lagrangian for this system reads

\[
{\cal{L}} = \frac{1}{2} m R^2 \dot{\theta}^2 +\frac{1}{2}m
\omega^2 R^2 \sin^2{\theta}+mgR\cos{\theta} \;\;.
\]

\begin{figure}[bt]
\centerline{\epsfig{file=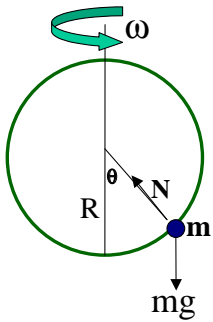,width=3.5cm,angle=0}}
\caption{\label{Fig. 1.} Mechanical rotating system.}
\end{figure}

\noindent Last equation can be rewritten as ${\cal{L}}
= T-U$, with $T$ being the kinetic energy and $U$ an effective potential
energy that includes the effects of gravity and the rotation of the
system. Introducing a dimensionless form one finds that

\[
{\cal U} =
\frac{U}{mgR}=-\cos{\theta}-\frac{1}{2}\frac{R\omega^2}{g}\sin^2{\theta}
\;\;.
\]

Therefore the analogy between this mechanical system and the
magnetic particle describes by Eq. 1 is clear. The centripetal
acceleration, $R\omega^2$, plays the role of the magnetic
anisotropy whereas the gravity field is equivalent to the external
magnetic field. Obviously, the downwards component can be further
increased by means of the application of an additional force, and
in such case the total force F divided by the mass of the particle
represents the applied field. Notice that achieving the anisotropy
field is equivalent to apply a downwards force which completely
compensates the vertical component of the normal force on the
particle.

When we analyze this mechanical system, a second order
phase transition appears. Analogously, analyzing the magnetization
along the field as a function of the parameter $H_a/H$, a second
order phase transition is clearly observed. In the latter case, the
order parameter is the angle $\theta$. Figure 2 shows the energy
as a function of $\theta$ for two different values of $H_a/H$.
Clearly, there is a smooth transition between one minimum in the
energy landscape, for $H_a/H < 1$ , to two minima, for $H_a/H >
1$. The transition occurs for an external field strong enough to
make the equilibrium state a magnetized one along the field. On
the other hand, when the anisotropy field is stronger, there are
two minima. They correspond to the states with non zero
magnetization component along the uniaxial anisotropy axis
together with the magnetization component along the external
field. Then, there is a second order phase transition for $H_a/H =
1$, in analogy with the well-known SW predictions.

\begin{figure}[bt]
\centerline{\epsfig{file=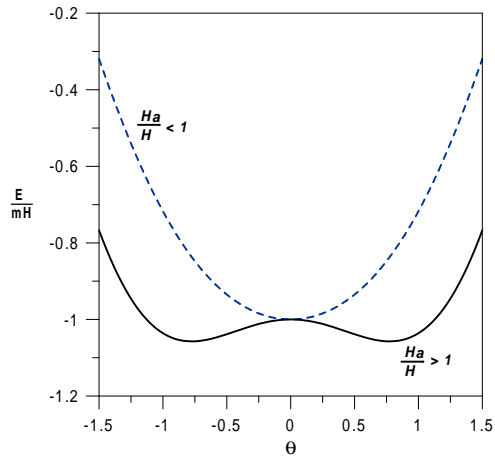,width=9cm,angle=0}}
\caption{\label{Fig. 2} Reduced energy landscape for a magnetic
system with anisotropy axis perpendicular to the external field.
The energy is plotted against the angle $\theta$ between the
magnetization of the system and the external field. We illustrate
two representative regimes, as a function of the ratio $H_a/H$.
The broken curve illustrates results for $H_a/H= 0.5$ and the
continuous line for $H_a/H=1.4$.}
\end{figure}

\begin{figure}[bt]
\centerline{\epsfig{file=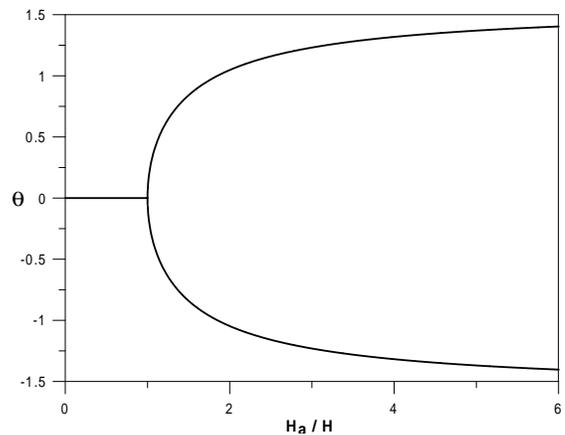,width=8cm,angle=0}}
\caption{\label{Fig. 3} Equilibrium angle, $\theta$,  between
magnetization and external field,as a function of the ratio
$H_a/H$.}
\end{figure}

\noindent Figure 3 illustrates the order parameter  $\theta$ as a
function of the ratio $H_a/H$. The second order phase transition
is clearly evidenced.  This figure also illustrates the
position-frequency phase diagram for the mechanical analog
depicted in figure 1.

\noindent Now we investigate the thermodynamic of this  particular
system in the canonical ensemble. The partition function $Z_{2D}$
reads:

\[
Z_{2D} = \int_0^{2\pi} d\theta \exp{\left(\frac{m H \cos{\theta} +
\frac{1}{2} m H_a \sin^2 \theta}{kT}\right)} \;\;.
\]

\noindent This integral can be analytically solved and the
result can be written as a series in Bessel functions of integer
argument.

\[
Z_{2D} = 2 \pi \sum_{n=0}^{\infty} {{2\,n}\choose n} \left(
\frac{H_a}{4 H} \right) ^n I_n (\frac{mH}{kT})  \; \;. \]

\noindent By making $H_a = 0$ we recover the ideal case of a
paramagnetic system in 2D. Hereafter thermodynamics follows and it
can be computed analytically. Magnetization and susceptibility of
the particle along the field are respectively given by

\[
<m> \; = \; kT \frac{\partial \ln Z_{2D}}{\partial H} \; \; \; \;
; \;\;\;\; \chi = \frac{\partial<m>}{\partial H} \; .
\]

\noindent These formula are evaluated using  the recurrence
relations for the first derivative of the Bessel functions.

Figure 4 illustrates the magnetization along the
external field as a function of temperature for a system of non
interacting particles with $m=1000\mu_B$ each and external field
$H=0.1kOe$. At $T=29.6$K the magnetization exhibits a maximum
 for  $H_a/H =30$. This behavior is clearly
understood by using the mechanical analog, where at fixed rotation
speed $\omega$ and zero kinetic energy, the equilibrium angle
$\theta$ is larger than the value obtained by a small increasing
of kinetic energy in the particle, due to the asymmetry of the
energy landscape around a minimum. Further increasing in kinetic
energy makes the angle increases up to an asymptotic final value
of $\theta=\pi/2$. In our magnetic system, larger kinetic energy
corresponds to higher temperatures. Therefore, in equilibrium and
for $H_a/H > 1$, the magnetization exhibits a maximum at a
temperature $T$ such that $kT=0.152(H_a/H-1)mH$

Figure 5 displays the magnetic susceptibility along the
field  direction as a function of temperature. The broken line
depicts results for $H_a/H=0.9$ and the continuous line for
$H_a/H=30$. In the region of the double well  potential, $H_a/H > 1
$, one clearly observes a maximum in both magnetization and
susceptibility curves.

\noindent The maximum of the susceptibility for $H_a/H << 1 $
approximately occurs at a temperature $T_{max}$ given by
$mH/kT_{max}=1.33$, which corresponds to the value for a
paramagnetic particle in 2D. For $H_a/H = 30$, the maximum occurs
at $mH/kT_{max}=1.0$.

\begin{figure}[bt]
\centerline{\epsfig{file=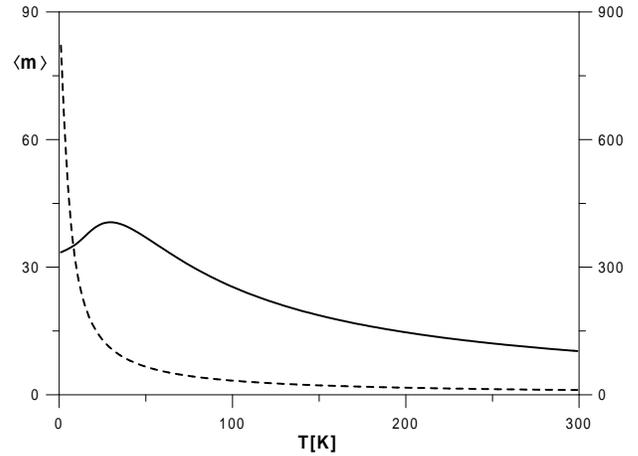,width=9cm,angle=0}}
\caption{\label{Fig. 4} Magnetization along the external field
axis, in units of $\mu_B$, as a function of temperature for a
system of non interacting particles with $m=1000 \mu_B$ each, and
external field $H=0.1kOe$. The continuous line, with magnetization
scale at the left, illustrates results for $H_a=3kOe$. The broken
line, with magnetization scale at the right, depicts results for
$H_a=0.09kOe$.}
\end{figure}

\begin{figure}[bt]
\centerline{\epsfig{file=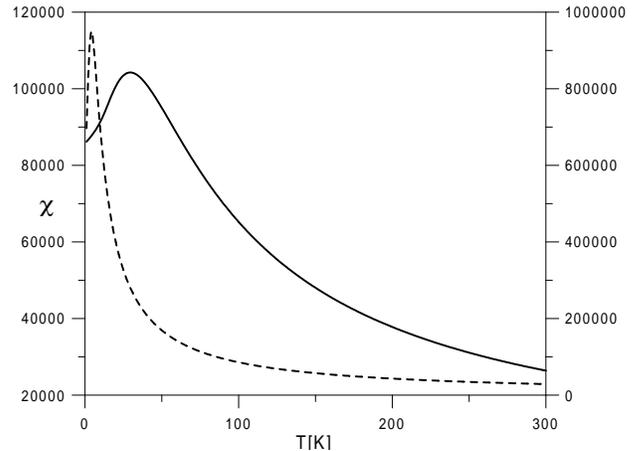,width=9cm,angle=0}}
\caption{\label{Fig.5} Magnetic susceptibility along the external
field, in units of $\mu_B /kOe$, as a function of temperature for
a system of non interacting particles with $m=1000\mu_B$ each, and
external field $H=0.1kOe$. The continuous line, with magnetization
scale at the left, illustrates results for $H_a=3kOe$. The broken
line, with magnetization scale at the right, depicts results for
$H_a=0.09kOe$.}
\end{figure}

\noindent Finally, in order  to explore the role of anisotropy,
Figure 6 illustrates the inverse susceptibility for different
angles $\alpha$ between anisotropy axis and external magnetic
field.

\noindent For a parallel orientation between anisotropy axis and
magnetic field we observe a susceptibility curve resembling a
ferromagnetic system.

\begin{figure}[bt]
\centerline{\epsfig{file=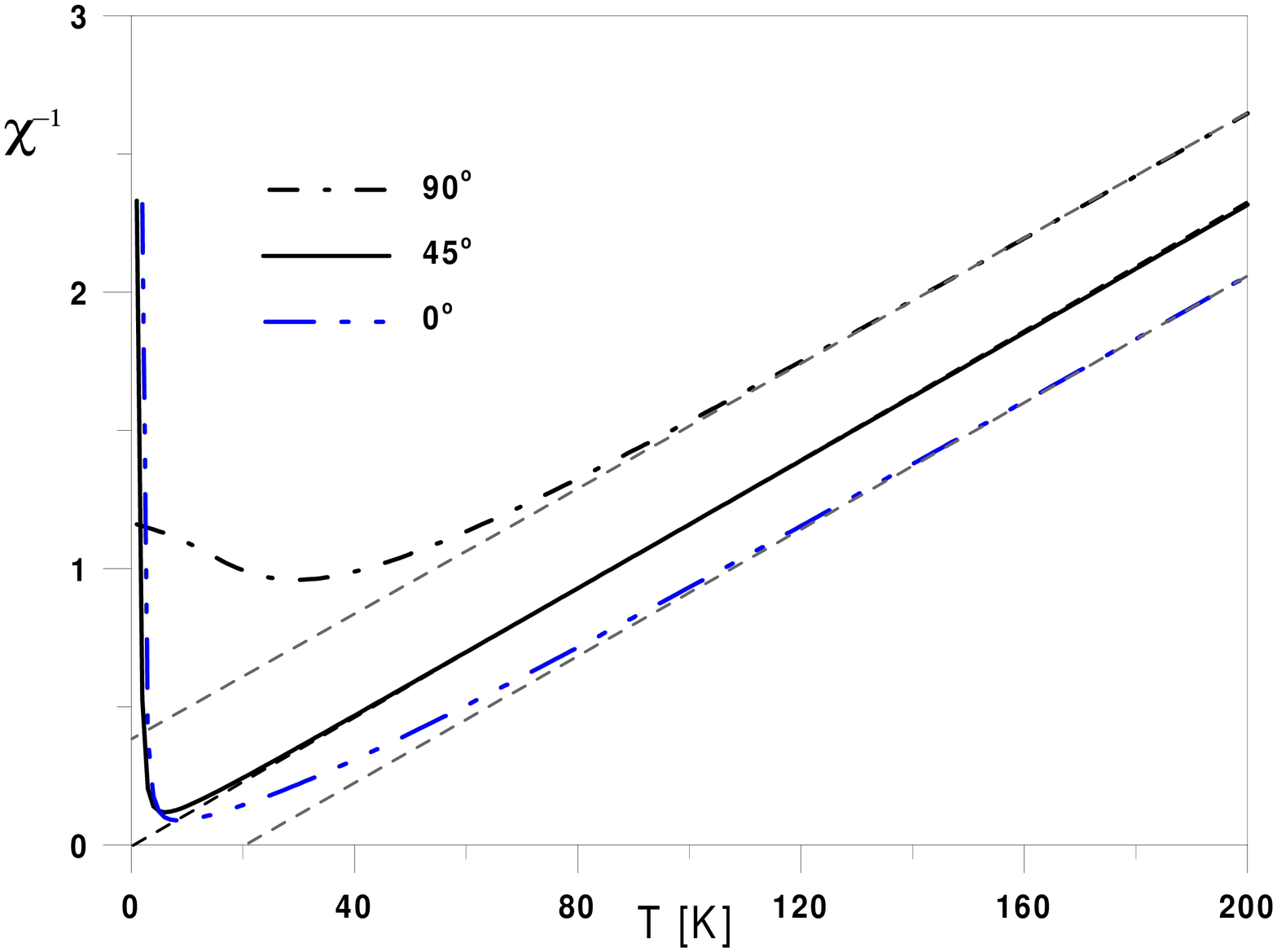,width=9cm,angle=0}}
\caption{\label{Fig.6} Inverse Magnetic susceptibilities in units
of $H/m$ along the external field as a function of temperature for
a system of non interacting particles with $m=1000\mu_B$ each,
external field $H=0.1kOe$ and $H_a=3kOe$. We show three curves
depending on the angle $\alpha$ between anisotropy axis and
external field as indicated inside the figure. The dashed straight
lines are high temperature extrapolations.}
\end{figure}

However, for perpendicular orientation between magnetic field and
anisotropy axis, we observe a typical Curie-Weiss
antiferromagnetic-like behavior, where by extrapolating from
higher temperatures, the zero crossing in the temperature axis
occurs at negative values.  When $\alpha=45^o$, a simple Curie law
is observed. Then, the extrapolated transition temperature changes
continuously from a positive value, for parallel orientation, to a
negative one,  when the anisotropy axis is perpendicular to the
external field. As a function of the relative angle between the
anisotropy axis and external field, the net effect of the
anisotropy field resembles a second order transition from a
ferromagnetic to an antiferromagnetic coupling.

In conclusion, a system of non-interacting  magnetic particles
with parallel anisotropy axes under the presence of an external
magnetic field applied perpendicular to the anisotropy axes
exhibits (at $T = 0K$) a second order phase transition as a
function of the ratio between external field to the anisotropy
field. The order parameter being the magnetization of the system.
Such system is isomorph to a mechanical system consisting in a
particle moving without friction in a circle rotating about its
vertical diameter. At finite temperature the thermodynamics of the
system can be solved analytically and the magnetization along
external magnetic field shows a maximum at a finite temperature,
contrary to that of a paramagnetic particle. The inverse
susceptibility of a system of non interacting magnetic particles
as a function of temperature demonstrates that uniaxial anisotropy
acts similarly to effective ferromagnetic or antiferromagnetic
coupling, depending on the angle between magnetic field and
anisotropy axis. The presented results can be useful in the
interpretation of magnetic data obtained in nanocrystalline
materials, specially novel artificially grown patterned media for
high density magnetic recording, which, depending on the aspect
ratio, can display a strong 2D character.

\section*{Acknowledgment}
This research received financial support from FONDECYT under
grants 1990812 and 1010127.  A bilateral project Vitae/Fundacion
Andes and  Millennium Science Nucleus "Condensed Matter Physics"
P99-135F are also acknowledged


\begin{references}

\bibitem{Jamet01}
M. Jamet, W. Wernsdorfer, C. Thirion, D. Mailly, V. Dupuis, P. Mélinon, A.
Pérez, Phys. Rev. Lett. 86 (20), 4676-4679 (2001).

\bibitem{DanDahlberg99}
E. Dan Dahlberg and R. Proksch, J. Magn. Magn. Mater. 200, 720-728 (1999)
and references therein.

\bibitem{Wernsdorfer00}
W. Wernsdorfer, D. Mailly and A. Benoit, J. Appl. Phys. 87 (9), 5094-5096
(2000) and references therein.

\bibitem{Rothman01}
J. Rothman et al. Phys. Rev. Lett. 86, 1098 (2001).

\bibitem{stoner48}
E.C. Stoner, and E.P. Wohlfarth, Phil. Trans. Roy. Soc. {\bf A 240}, 599
(1948); Reprinted by IEEE Trans. Magn.{\bf 27} (4), 3475 (1991).

\bibitem{Bonet99}
E. Bonet, W. Wernsdorfer, B. Barbara, A. Benoit, D. Mailly and A.
Thiaville, Phys. Rev. Lett. 83 (20) 4188-4191 (1999).

\bibitem{Wernsdorfer97}
W. Wernsdorfer, E. Bonet Orozco, K. Hasselbach, A. Benoit, B. Barabara, N.
Demoncy, A. Loiseau, H. Pascard, D. Mailly, Phys. Rev. Lett. 78 (9)
1791-1794 (1997).

\bibitem{garciaotero98}
J. Garcia-Otero, A.J. Garcia-Bastida and J. Rivas, J. Magn. Magn. Mater.
{\bf 189}, 377 (1998).

\bibitem{Dimitrov96}
D.A. Dimitrov and G.M. Wysin, Phys. Rev. B {\bf 54}, 9237 (1996).

\bibitem{Neel49}
L. N\'{e}el, Ann. Geophys. 5, 99 (1949).

\bibitem{Brown63}
W.F. Brown, Phys. rev. 130, 1677 (1963).

\bibitem{Igarashi00}
M. Igarashi, F. Akagi, K. Yoshida and Y. Nakatani, IEEE Trans.
Magn. 36 (5) 2459-2461 (2000).

\bibitem{Cregg99}
P. J. Cregg and L. Bessais, J. Magn. Magn. Mater. 202, 554-564
(1999).

\bibitem{Respaud99}
M. Respaud, J. Appl. Phys. 86 (1), 556-561 (1999).

\bibitem{Pfannes00}
H.-D. Pfannes, A. Mijovilovich, R. Magalh\~{a}es-Paniago and R.
Paniago, Phys. Rev. B 62 (5), 3372-3380 (2000).

\bibitem{Allia99}
P.Allia, M. Coisson, M.Knobel, P. Tiberto and F.Vinai, Phys. Rev. B {\bf 60} (17) 12207-12218 (1999).

\bibitem{Anderson97}
J.-O. Anderson, C. Djuberg, T. Jonsson, P. Svedlindh and P.
Norblad, PRB 56 (21), 13983-13988 (1997).

\bibitem{Allia01}
P.Allia, M. Coisson, M.Knobel, P.Tiberto and
F.Vinai, M.A. Novak and W.C. Nunes, Phys. Rev. B (in press).

\bibitem{Denardin01}
J.C. Denardin, A.L. Brandl, M. Knobel, P. Panissod, X.X. Zhang, A.B.
Pakhomov, H. Nie, unpublished.

\end{references}
\end{document}